\documentstyle[aps,multicol,epsf]{revtex}
\begin{document}
%\draft
%\preprint{}
\title{Casimir Dispersion Forces and Orientational Pairwise Additivity}
\author{Ramin Golestanian}
\address{
Institute for Theoretical Physics, University of California,
Santa Barbara, CA 93106 \\
and Institute for Advanced Studies in Basic Sciences, Zanjan 45195-159, Iran}
\date{\today}
\maketitle
\begin{abstract}

A path integral formulation is used to study the fluctuation--induced interactions 
between manifolds of arbitrary shape at large separations. 
It is shown that the form of the interactions crucially depends
on the choice of the boundary condition. In particular, whether or not the
Casimir interaction is pairwise additive is shown to depend on whether
the ``metallic'' boundary condition corresponds to a ``grounded'' or an ``isolated''
manifold.

\end{abstract}
\begin{multicols}{2}%\narrowtext

\section{Introduction and summary}	\label{sIntro}

External objects that are immersed in a fluctuating medium, and
modify the fluctuations in their vicinity, experience induced
interactions with one another \cite{Casimir,Dzya,mostepa,Krech,KGRMP}.
These interactions are most often independent of the structural
details, and are in turn highly sensitive to the
geometry of the objects and their mutual arrangements while 
immersed in the medium.

The strong dependence of these interactions on the shape of the objects
raise the issue of {\em pairwise additivity}: Is it possible to express the fluctuation--induced interaction between two extended bodies as the sum
of a pair potential, or the interaction between several bodies as the sum
of two-body interactions? 

It is well known that a pairwise summation of the van der Waals interaction
gives the correct power law for the Casimir energy \cite{mostepa}. Let us
take a pair potential of the form $-A/r^n$, with $n=6$ for the thermal case
and $n=7$ for the quantum case \cite{Dzya}, and $A$ being a constant 
to be determined. If one tries to fix the coefficient by summing the pair potential
over two bodies and equating the result to the expression for the Casimir interaction
between the bodies, one finds out that a different coefficient is needed for every
geometry. 

To understand this, one should note that the van der Waals interaction
is due to dipolar fluctuations. When two extended bodies are at a close separation,
one can show that the fluctuations of all the multipoles in fact contribute comparably
to the Casimir energy, and thus summation of the contribution due to the dipolar
fluctuations cannot by itself account for the interaction \cite{HaLiu1}.
When the bodies are at large separations (larger than their typical sizes), 
the contribution due to higher multipoles are in fact systematically weaker. 
However, there is still a discrepancy between the sum of a (van der Waals)
pair potential, and the contribution of the dipolar fluctuations to the Casimir
energy. In the spirit of a (second order) perturbation theory, the correct way of calculating the dipolar Casimir energy is to consider the pairwise sum of the 
dipole-dipole interactions over the two bodies, and then square it and take 
the average. This is clearly in contrast with the pairwise summation of the 
van der Waals interaction, which corresponds to taking the square of the
local dipolar fluctuations and averaging, and then summing over the two bodies.

The same picture can help us answer the second question. Many-body interactions
can be expressed as the sum of many-body interactions of the 
multipoles of different bodies in the medium. When extended bodies are at close
separations, and all the multipoles have comparable contributions, many-body
interactions of nontrivial forms result \cite{LiK,HaLiu2,Pod}. On the other hand, for bodies at large separations the leading order contribution comes from the sum of two-body interactions of the lowest nonvanishing multipole \cite{GGK}.

In this article, we study the issue of orientational pairwise additivity \cite{GGK},
which is to determine whether the orientational dependence of the interactions could be obtained from the summation of a pair potential.
A path integral formulation is used to study the fluctuation--induced interactions between manifolds of arbitrary shape at large separations, in the context of a multipole 
expansion. It is shown that the form of the interaction crucially depends
on whether the manifolds are {\em grounded} or {\em isolated} in an electrostatic analogy. 
In the grounded case, the manifolds are connected to a {\em charge reservoir} to maintain
a constant {\em potential}, and thus the leading fluctuations are {\em monopolar}.
Isolated manifolds, however, are constrained to have fixed overall charges, and can
only undergo {\em dipolar} fluctuations. The leading interaction between grounded
manifolds is found to be of the form $({\rm monopole-monopole})^2$, and is independent
of their shapes and orientations. The leading shape dependent term comes from the  
$({\rm monopole-dipole})^2$ term, which gives rise to orientational dependencies
that are pairwise additive. The interaction between isolated manifolds, however, is
dominated by the  $({\rm dipole-dipole})^2$ term to the leading order, which is
{\em not} pairwise additive.

The rest of the paper is organized as follows. In Sec. \ref{sPath}, the path integral
formulation is developed and general expressions are derived for the fluctuation--induced
interactions for different types of boundary conditions. 
In Sec. \ref{sAppl}, the interactions are examined for the specific examples of
symmetric objects such as spheres, and also highly asymmetric objects such as rods and disks, where the above features can be manifestly understood. 
Critical fluids are examined in Sec. \ref{sCritical}, as a special case, and
a conclusion follows in Sec. \ref{sDiscuss}.

\section{Path Integral Formulation}	\label{sPath}
Consider a $d$-dimensional medium, in which a field $\phi$ is undergoing scale-free (massless)
thermal fluctuations, described by the Hamiltonian
\begin{equation}	\label{H}
{\cal H}[\phi]={K \over 2} \int d^d {\bf x} (\nabla \phi)^2.
\end{equation}
The field could represent a component of the electromagnetic field in a dielectric medium
\cite{mostepa}, the electrostatic potential in charged fluids at very low salt concentrations 
\cite{KGRMP}, an order parameter field for a critical binary mixture or a magnetic system \cite{dGF}, or a massless Goldstone mode arising from a continuous symmetry breaking \cite{LiK}. For simplicity, in what follows we are going to think in the context of charged fluids, and later on comment on the specific effects in the case of critical fluids.

Let us assume that there are $n$ manifolds immersed in the medium, denoted by 
$M_\alpha$ ($\alpha=1, \cdots, n$), which modify the fluctuations. In the electrostatic
context, one can view each manifold as a {\em conductor} that requires a constant value
for the potential field in the whole volume that it encloses.
A restricted partition function, which requires a value of $\phi_\alpha$ for the 
potential field on the $\alpha$th manifold, can then be written as
\begin{equation}	\label{Z1}
{\cal Z}[\phi_\alpha]=\int {\cal D} \phi({\bf x}) \prod_{\alpha=1}^{n} 
\delta\left\{\phi |_{M_\alpha}-\phi_\alpha \right\} \; {\rm e}^{-{\cal H}[\phi]}.
\end{equation}
Following Ref. \cite{LiK}, the functional delta functions can next be represented
by introducing the Lagrange multiplier fields $\rho_\alpha({\bf x})$, as
\end{multicols}
\begin{eqnarray}	\label{Z2}
{\cal Z}[\phi_\alpha]&=&\int {\cal D} \phi({\bf x}) \prod_{\alpha=1}^{n}  \int_{M_\alpha} {\cal D}\rho_\alpha({\bf x}) 
\; \exp\left\{-{K \over 2} \int d^d {\bf x} (\nabla \phi)^2 
+ i \sum_{\alpha} \int d^d {\bf x} \rho_\alpha({\bf x}) \left[\phi({\bf x})-\phi_\alpha\right] \right\} \\
&=& {\cal Z}_0 \times \prod_{\alpha=1}^{n}  \int_{M_\alpha} {\cal D}\rho_\alpha({\bf x})
\;\exp\left\{-{1 \over 2 K} \sum_{\alpha,\beta} \int d^d {\bf x} d^d {\bf x}' \rho_\alpha({\bf x}) G({\bf x}-{\bf x}') 
\rho_\beta({\bf x}')-i \sum_{\alpha} \phi_\alpha \int d^d {\bf x} \rho_\alpha({\bf x}) \right\}, \nonumber
\end{eqnarray}
\begin{multicols}{2}\noindent
in which 
\begin{equation}	\label{G}
G({\bf x}-{\bf x}')=(-\nabla^2)^{-1}_{{\bf x},{\bf x}'}={1 \over S_d (d-2) |{\bf x}-{\bf x}'|^{d-2}},
\end{equation}
with $S_d={2 \pi^{d/2} \over \Gamma(d/2)}$  (the surface area of the $d$-dimensional sphere), ${\cal Z}_0$
is the free partition function, and $\int_{M_\alpha} {\cal D}\rho_\alpha({\bf x})$ implies a functional 
integration only in the region enclosed by $M_\alpha$. (In other words, the Lagrange multiplier field 
$\rho_\alpha({\bf x})$ is nonzero only within the volume of $M_\alpha$.) Note that one should view 
the $\rho_\alpha({\bf x})$ fields as fluctuating charge density fields, and Eq.(\ref{Z2}) as the partition 
function of a set of interacting Coulomb plasmas, in the electrostatic context \cite{GBP}.

The fluctuation--induced interactions between the conductors can now be inferred from the
above partition function. However, it is important to specify the boundary conditions for
the conductors. One possibility is that the conductors are grounded, that is to say they
are maintained at a constant fixed potential by being in contact with a large reservoir of 
charges; a so-called ``ground.'' In this case, the free energy of the system is obtained as
\begin{equation}	\label{bFG}
F_{\rm gr}=- k_{\rm B} T \ln {\cal Z}[\phi_\alpha=0].
\end{equation}

The other possibility is that the conductors are made isolated, and maintain constant
amounts of net charges, which we assume to be zero. In this case, the 
potential field at the conductors can take any value to help maintain the neutrality, and thus the
free energy is obtained as
\begin{equation}	\label{bFI}
F_{\rm is}=- k_{\rm B} T \ln \left(\int_{-\infty}^{+\infty} \prod_{\alpha} d \phi_\alpha {\cal Z}[\phi_\alpha]\right).
\end{equation}

To further proceed, we focus on the situation in which the manifolds are far from each other,
namely, they are at separations much larger than their typical sizes. In this case, we can perform a multipole expansion for the charge density distribution. For example, the Coulomb interaction between the $\alpha$th and the $\beta$th conductors can be written as ($\alpha \neq \beta$)
\end{multicols}
\begin{eqnarray}	
h_{\alpha \beta}&=&\int d^d {\bf x} d^d {\bf x}' {\rho_\alpha({\bf x}) \rho_\beta({\bf x}') 
\over S_d (d-2) |{\bf x}-{\bf x}'+{\bf R}_{\alpha \beta}|^{d-2}}, \nonumber \\
&=&{Q_\alpha Q_\beta \over S_d (d-2) R_{\alpha \beta}^{d-2}}
-\left(Q_\beta {\bf P}_\alpha \cdot {\hat {\bf R}}_{\alpha \beta} 
-Q_\alpha {\bf P}_\beta \cdot {\hat {\bf R}}_{\alpha \beta}\over S_d R_{\alpha \beta}^{d-1} \right)
+\left( {\bf P}_\alpha \cdot {\bf P}_\beta -d  {\bf P}_\alpha \cdot {\hat {\bf R}}_{\alpha \beta}
{\bf P}_\beta \cdot {\hat {\bf R}}_{\alpha \beta} \over S_d R_{\alpha \beta}^d\right)+ \cdots,\label{hab}
\end{eqnarray}
\begin{multicols}{2}\noindent
in which $R_{\alpha \beta}$ is the distance between the two conductors, and the multipoles
are defined as
\begin{eqnarray}	\label{Qdef}
Q_\alpha&=& \left.\int d^d {\bf x}  \rho_\alpha({\bf x})=\tilde{\rho}_\alpha({\bf k})\right|_{{\bf k}={\bf 0}},\\
P_{\alpha,i}&=& \int d^d {\bf x} x_i \rho_\alpha({\bf x})=\left.{1 \over i} {\partial \over \partial k_i}
\tilde{\rho}_\alpha({\bf k})\right|_{{\bf k}={\bf 0}},	\label{Pdef}
\end{eqnarray}
with
\begin{equation}	\label{rhotilde}
\tilde{\rho}_\alpha({\bf k})=\int d^d {\bf x} \rho_\alpha({\bf x}) {\rm e}^{i {\bf k} \cdot {\bf x}}.
\end{equation}

We also need to make a similar multipole expansion for the self energy terms at each manifold
\begin{eqnarray}
h_{\alpha \alpha}&=&\int d^d {\bf x} d^d {\bf x}' {\rho_\alpha({\bf x}) \rho_\alpha({\bf x}') 
\over S_d (d-2) |{\bf x}-{\bf x}'|^{d-2}}, \nonumber \\
&=& \int {d^d {\bf k} \over (2 \pi)^d} {1 \over k^2}\tilde{\rho}_\alpha({\bf k})\tilde{\rho}_\alpha(-{\bf k}).\label{haa}
\end{eqnarray}
We can introduce the multipoles, using the Taylor expansion of the charge density in Fourier space
\begin{eqnarray}
\tilde{\rho}_\alpha({\bf k})&=&\tilde{\rho}_\alpha({\bf 0})
+\left.{1 \over i} {\partial \tilde{\rho}_\alpha({\bf k})\over \partial k_i}\right|_{{\bf k}={\bf 0}} i k_i+\cdots,\nonumber \\
&=& Q_\alpha+P_{\alpha,i} \times i k_i+\cdots,\label{Taylor}
\end{eqnarray}
The above expansion can be formally viewed as an expansion in powers of $k L_\alpha$, where $L_\alpha$ is a typical size 
of the manifold. The expansion is thus convergent only for sufficiently small values of $k$, corresponding to length
scales larger than the size of the manifolds. Since the self energy integral in Eq.(\ref{haa}) involves contributions 
from higher wavevectors, a multipole expansion for the self energy will be divergent.
However, the expansion in Eq.(\ref{Taylor}) indicates that all the information concerning the first few multipoles
of the charge distribution is already contained in the low $k$ behavior of the function $\tilde{\rho}_\alpha({\bf k})$.
Since we are only interested in the dependence of the partition function Eq.(\ref{Z2}) on the distances 
$R_{\alpha \beta}$, all we need to know about the self energy is its dependence on the first few multipoles, which
is in fact well behaved.

Let us denote the domain of convergence for the expansion in Eq.(\ref{Taylor}) in $k$-space by ${\cal D}_\alpha$.
This domain contains the origin, and its shape is determined by the geometry of the conductor. Loosely speaking,
its size in each direction is set by the inverse of the size of the conductor in that direction. Now we can restrict
the $k$-integral in the self energy only to this domain, and neglect the contribution from the outside of 
${\cal D}_\alpha$, because all the dependence on the first few multipoles is included in the domain ${\cal D}_\alpha$.
We thus have
\begin{eqnarray}
h_{\alpha \alpha}&=&\int_{{\cal D}_\alpha} {d^d {\bf k} \over (2 \pi)^d} {1 \over k^2}
\tilde{\rho}_\alpha({\bf k})\tilde{\rho}_\alpha(-{\bf k})+\cdots,\nonumber \\
&=&\gamma_\alpha Q_\alpha^2+\gamma_{\alpha,ij} P_{\alpha,i} P_{\alpha,j}+\cdots,	\label{gamma}
\end{eqnarray}
in which
\begin{eqnarray}
\gamma_\alpha &=&\int_{{\cal D}_\alpha} {d^d {\bf k} \over (2 \pi)^d} {1 \over k^2},\label{gam} \\
\gamma_{\alpha,ij}&=&\int_{{\cal D}_\alpha} {d^d {\bf k} \over (2 \pi)^d} {k_i k_j \over k^2},\label{gamij}
\end{eqnarray}
and so forth.

Putting all the pieces together, the $R_{\alpha \beta}$-dependent part of the partition function can be written as 
\end{multicols}
\begin{eqnarray}	\label{Zphi}
{\cal Z}[\phi_\alpha]&=& \int \prod_{\alpha} d Q_\alpha d {\bf P}_{\alpha} \cdots 
{\rm e}^{-i \sum_{\alpha} \phi_\alpha Q_\alpha} \times 
\exp\left\{-{1 \over 2 K} \sum_{\alpha}\left[\gamma_\alpha Q_\alpha^2+\gamma_{\alpha,ij} P_{\alpha,i} P_{\alpha,j}+\cdots\right]\right. \\
&&\left.-{1 \over 2 K} \sum_{\alpha\neq\beta}\left[{Q_\alpha Q_\beta \over S_d (d-2) R_{\alpha \beta}^{d-2}}
-{{\hat R}_{\alpha \beta,i} (Q_\beta P_{\alpha,i} -Q_\alpha P_{\beta,i}) \over S_d R_{\alpha \beta}^{d-1}}
+{(\delta_{ij} -d {\hat R}_{\alpha \beta,i}{\hat R}_{\alpha \beta,j}) P_{\alpha,i} P_{\beta,j} 
\over S_d R_{\alpha \beta}^d}+ \cdots\right]\right\}.\nonumber
\end{eqnarray}
\begin{multicols}{2}
Note that we have neglected a Jacobian in changing the measure of integration. However, since the transformation from the charge density distribution to the multipole description is linear, one can show that the Jacobian is just an uninteresting constant.

Finally, using Eqs.(\ref{bFG}) and (\ref{Zphi}), the interaction 
free energy for grounded manifolds can be obtained as 
\end{multicols}
\begin{equation}	\label{FG}
F_{\rm gr}=- {k_{\rm B} T \over 4 S_d^2} \sum_{\alpha\neq\beta} 
\left[{\gamma_\alpha^{-1} \gamma_\beta^{-1}\over (d-2)^2 R_{\alpha \beta}^{2(d-2)}}
+{(\gamma_\alpha^{-1} \gamma_{\beta,ij}^{-1}+\gamma_{\alpha,ij}^{-1} \gamma_\beta^{-1})
{\hat R}_{\alpha \beta,i}{\hat R}_{\alpha \beta,j}\over R_{\alpha \beta}^{2(d-1)}}
\right]+O(1/R^{2d}).
\end{equation}
The first term in Eq.(\ref{FG}) is a squared monopole--monopole interaction, and is independent of the relative orientations of the conductors in space. The second term,
on the other hand, has the form of a squared monopole--dipole interaction, and does
depend on the orientations through an effective dipole--dipole interaction, which is 
pairwise additive. 

Similarly, for isolated manifolds, Eqs.(\ref{bFI}) and (\ref{Zphi}) yield
the interaction as
\begin{equation}	\label{FI}
F_{\rm is}=- {k_{\rm B} T \over 4 S_d^2} \sum_{\alpha\neq\beta} 
{\gamma_{\alpha,ik}^{-1} \gamma_{\beta,jl}^{-1}\over R_{\alpha \beta}^{2d}}
(\delta_{ij} -d {\hat R}_{\alpha \beta,i}{\hat R}_{\alpha \beta,j})
(\delta_{kl} -d {\hat R}_{\alpha \beta,k}{\hat R}_{\alpha \beta,l})+O(1/R^{2d+2}).
\end{equation}
\begin{multicols}{2}\noindent
Note that the leading term in Eq.(\ref{FI}) is a squared dipole--dipole interaction,
and thus it is not orientationally pairwise additive.

\section{Application to Specific Geometries}	\label{sAppl}

The multipole expansion allowed us to calculate the general forms of the fluctuation--induced
interactions between manifolds of arbitrary shape and with arbitrary orientations with respect to one another, for the two cases of isolated and grounded boundary conditions. All the specific 
informations about the shapes and the orientations of the manifolds are encoded in the 
$\gamma$-tensors defined above. These informations are in fact of three kinds: (i) the overall
magnitude of the tensors which are set by the typical sizes of the manifolds, (ii) the orientational dependencies which make up the tensorial structure, and are dictated by the
structure of the symmetry axes or ``the principal axes'' of the manifolds, and (iii) overall
numerical prefactors of order unity. In this section, we try to use symmetry arguments
to determine the $\gamma$-tensors for some simple geometries within the numerical prefactors,
without actually specifying the exact shape of the integration domain ${\cal D}$. The final piece of information, which is the numerical prefactor, appears to be very sensitive to the
exact geometry of the manifold (and thus to that of ${\cal D}$), and can in general be calculated using the techniques developed in Ref. \cite{GGK}.

\subsection{Two Spheres}	\label{sSphere}

The $\gamma$-tensors for a sphere of radius $L$ can be easily estimated
using symmetry: $\gamma_s \sim \int_0^{1/L} k^{d-1} d k/k^2 \sim 1/L^{d-2}$
and $\gamma_{s,ij} \sim \delta_{ij}\int_0^{1/L} k^{d-1} d k \sim \delta_{ij}/L^{d}$.
Using Eqs.(\ref{FG}) and (\ref{FI}), the interaction between a sphere of radius
$L_1$, and another sphere of radius $L_2$ which is at a distance $R$, reads
\begin{equation}
F_{\rm gr}^{\rm sph} \sim - k_{\rm B} T \times {L_1^{d-2} L_2^{d-2}\over
R^{2(d-2)}},\label{GRsph}
\end{equation}
for the grounded case, and
\begin{equation}
F_{\rm is}^{\rm sph} \sim - k_{\rm B} T \times {L_1^{d} L_2^{d}\over
R^{2d}},\label{ISsph}
\end{equation}
for the isolated case, with no orientational dependence due to symmetry.

\subsection{Two Rods}	\label{sRod}

The calculation of the $\gamma$-tensors for a rod of length $L$ and 
thickness $a$ is more tricky. Using the cylindrical symmetry
one obtains: $\gamma_r \sim \int_0^{1/L}d k_z \int_0^{1/a} d k_{\perp}
k_{\perp}^{d-2}/(k_z^2+k_{\perp}^2) \sim 1/L^{d-2}$ for $d \leq 3$,
and $\sim 1/(L a^{d-3})$ for $d > 3$, where $z$-axis is parallel to the
director of the cylinder, and $\perp$ denotes the remaining directions
that are perpendicular to it.
The second rank tensor $\gamma_{r,ij}$ is diagonal with only two independent
components: $\gamma_{r,zz} \sim \int_0^{1/L}d k_z \int_0^{1/a} d k_{\perp}
k_{\perp}^{d-2} k_z^2/(k_z^2+k_{\perp}^2) \sim 1/L^{d}$ for $d \leq 3$,
and $\sim 1/(L^3 a^{d-3})$ for $d > 3$, and
$\gamma_{r,\perp \perp} \sim \int_0^{1/L}d k_z \int_0^{1/a} d k_{\perp}
k_{\perp}^{d}/(k_z^2+k_{\perp}^2) \sim 1/(L a^{d-1})$.
If the unit vector ${\hat {\bf d}}$ denotes the director of the rod, the
inverse second rank $\gamma$-tensor which appears in the expression
for the interaction can be written as $\gamma^{-1}_{r,ij} \sim L^d 
{\hat d}_i {\hat d}_j$ for $d \leq 3$, in the limit of small thickness.
Note that in this limit, the inverse $\gamma$-tensors are vanishing
for $d>3$, and thus rods do not interact in these high dimensions. 

Using Eqs.(\ref{FG}) and (\ref{FI}), the orientation dependent part of the
interaction between two rods of lengths $L_1$ and $L_2$, and directors
${\hat {\bf d}}_1$ and ${\hat {\bf d}}_{2}$, which are a 
distance $R$ apart, reads $(d \leq 3)$
\end{multicols}
\begin{equation}
F_{\rm gr}^{\rm rod} \sim - k_{\rm B} T \times {L_1^{d-1} L_2^{d-1}\over
R^{2(d-1)}} \times \left[{L_1 \over L_2} ({\hat {\bf d}}_1 \cdot
{\hat {\bf R}}_{12})^2+{L_2 \over L_1} ({\hat {\bf d}}_2 \cdot
{\hat {\bf R}}_{12})^2\right],\label{GRrod}
\end{equation}
for the grounded case, which is pairwise additive, and
\begin{equation}
F_{\rm is}^{\rm rod} \sim - k_{\rm B} T \times {L_1^{d} L_2^{d}\over
R^{2d}} \times \left[{\hat {\bf d}}_1 \cdot {\hat {\bf d}}_{2}- d
({\hat {\bf d}}_1 \cdot {\hat {\bf R}}_{12})
({\hat {\bf d}}_2 \cdot {\hat {\bf R}}_{12})\right]^2,\label{ISrod}
\end{equation}
\begin{multicols}{2}\noindent
for the isolated case, which has a squared dipolar form and is not 
pairwise additive.

\subsection{Two Disks}	\label{sDisk}

The $\gamma$-tensors for a disk of radius $L$ and thickness $a$ can be 
similarly calculated within numerical prefactors using symmetry. The zeroth
rank tensor can be calculated as: 
$\gamma_d \sim \int_0^{1/a}d k_z \int_0^{1/L} d k_{\perp}
k_{\perp}^{d-2}/(k_z^2+k_{\perp}^2) \sim 1/L^{d-2}$, where $z$-axis is normal
to the disk, and $\perp$ denotes the remaining directions in the subspace
of the disk.
The second rank tensor $\gamma_{d,ij}$ is diagonal with only two independent
components: $\gamma_{d,zz} \sim \int_0^{1/a} d k_z \int_0^{1/L} d k_{\perp}
k_{\perp}^{d-2} k_z^2/(k_z^2+k_{\perp}^2) \sim 1/(a L^{d-1})$, and
$\gamma_{d,\perp \perp} \sim \int_0^{1/a}d k_z \int_0^{1/L} d k_{\perp}
k_{\perp}^{d}/(k_z^2+k_{\perp}^2) \sim 1/L^{d}$.
If we denote the unit vector perpendicular to the disk by ${\hat {\bf n}}$,
the inverse second rank $\gamma$-tensor which appears in the expression
for the interaction can be written as $\gamma^{-1}_{d,ij} \sim L^d 
(\delta_{ij}-{\hat n}_i {\hat n}_j)$, in the limit of small thickness.

Using Eqs.(\ref{FG}) and (\ref{FI}), the orientation dependent part of the
interaction between a disk of radius $L_1$ and normal vector ${\hat {\bf n}}_1$,
and another one with radius $L_2$ and normal vector ${\hat {\bf n}}_{2}$ that is
a distance $R$ apart, reads
\end{multicols}
\begin{equation}
F_{\rm gr}^{\rm disk} \sim - k_{\rm B} T \times {L_1^{d-1} L_2^{d-1}\over
R^{2(d-1)}} \times \left\{{L_1 \over L_2}\left[1-({\hat {\bf n}}_1 \cdot
{\hat {\bf R}}_{12})^2\right]+{L_2 \over L_1} \left[1-({\hat {\bf n}}_2 \cdot
{\hat {\bf R}}_{12})^2\right]\right\},\label{GRdisk}
\end{equation}
for the grounded case, which is pairwise additive, and
\begin{eqnarray}	\label{ISdisk}
F_{\rm is}^{\rm disk} \sim - k_{\rm B} T \times {L_1^{d} L_2^{d}\over
R^{2d}} &\times& \left[d^2-d-2+({\hat {\bf n}}_1 \cdot {\hat {\bf n}}_{2})^2 
+(2 d-d^2) ({\hat {\bf n}}_1 \cdot {\hat {\bf R}}_{12})^2
+(2 d-d^2) ({\hat {\bf n}}_2 \cdot {\hat {\bf R}}_{12})^2 \right.\\
&&\left.-2 d ({\hat {\bf n}}_1 \cdot {\hat {\bf R}}_{12})
({\hat {\bf n}}_2 \cdot {\hat {\bf R}}_{12})
({\hat {\bf n}}_1 \cdot {\hat {\bf n}}_{2})
+d^2 ({\hat {\bf n}}_1 \cdot {\hat {\bf R}}_{12})^2
({\hat {\bf n}}_2 \cdot {\hat {\bf R}}_{12})^2\right] \nonumber,
\end{eqnarray}
\begin{multicols}{2}\noindent
for the isolated case, which has a squared dipolar form and is not 
pairwise additive.

\section{Critical Fluids}	\label{sCritical}

As mentioned above, interactions could be induced between objects that modify
thermal fluctuations of an order parameter field for a critical binary mixture 
or a magnetic system \cite{dGF}. In this case, two kinds of boundary conditions
are usually considered: (i) the {\it ordinary} boundary
condition that suppresses the order parameter at the boundary, and thus does not 
break its symmetry, and (ii) the {\it symmetry breaking} boundary condition,
which sets a nonvanishing value for the order parameter at the boundary.
Note that the ordinary boundary condition is the same as the grounded boundary
condition in the electrostatic terminology.

The fluctuations in a critical fluid are characterized by the universality class
of the system.
For the case when the fluid can be described by a Gaussian Hamiltonian as
given in Eq.(\ref{H}), all of the above results for the grounded manifolds hold
for the case of ordinary boundary condition. The interaction between manifolds
with symmetry breaking boundary conditions, where the value of the order parameter
is set to $\Phi_\alpha$ on the $\alpha$th manifold, is calculated as
\begin{equation}	\label{bFS}
F_{\rm sb}=- k_{\rm B} T \ln {\cal Z}[\phi_\alpha=\Phi_\alpha],
\end{equation}
where ${\cal Z}[\phi_\alpha]$ is given by Eq.(\ref{Zphi}). One obtains \cite{dG}
\begin{equation}	\label{FS}
F_{\rm sb}^{\rm Gauss}=- {k_{\rm B} T \over 2} \sum_{\alpha\neq\beta} 
\left[{K \Phi_\alpha \Phi_\beta \gamma_\alpha^{-1} \gamma_\beta^{-1}\over 
S_d (d-2) R_{\alpha \beta}^{d-2}}\right]+O(1/R^{2d-4}).
\end{equation}
It is important to note that this interaction is independent of the orientations
of the manifolds, and that the leading order orientation dependent term for the
symmetry breaking boundary condition is the same as the case of ordinary boundary condition,
and is given as in Eq.(\ref{FG}). This interaction is orientationally pairwise
additive.

For a nontrivial universality class, one should make use of more complicated
Hamiltonians with nonlinear terms. It is then possible to calculate the Casimir
energy expressions using field theoretical techniques \cite{Krech}. The interaction
between two spheres in an arbitrary critical system has in fact been calculated
exactly in Ref.\cite{BuEi} using conformal-invariance methods. The interaction
for the case of symmetry breaking boundary conditions (on both spheres) is obtained
as $1/R^{d-2+\eta}$\cite{betanu}, while for the case of ordinary boundary conditions 
it is found as $1/R^{2(d-1/\nu)}$, where $\eta$ and $\nu$ are critical exponents of 
the system \cite{BuEi}. One can easily check that for the case of Gaussian universality
class, where $\eta=0$ and $\nu=1/2$, they coincide with the results of Eq.(\ref{FS})
and (\ref{FG}). 

It is interesting to note that the power law for the symmetry breaking 
case is given by the two-point correlation function of the field (the spin-spin correlation
in magnetic terminology), while the one for the ordinary case is given by the
four-point correlation function (the energy-energy correlation) \cite{BuEi}.
Guided by this, one can think of an effective Gaussian Hamiltonian of the form
\begin{equation}	\label{Hcf}
{\cal H}_{\rm cf}[\phi]={K \over 2} \int {d^d {\bf q} \over (2 \pi)^d} q^{2-\eta} 
|\phi({\bf q})|^2,
\end{equation}
which yields a correct form for the two-point function, and calculate the 
fluctuation--induced interactions using the above methods \cite{LiK}.
However, although it yields a correct result for the symmetry breaking case
(almost by construction), it gives a corresponding form for the ordinary case
as $1/R^{2(d-2+\eta)}$ which is not correct. The reason is that the above effective
Gaussian Hamiltonian does not give a correct four-point correlation function. However, 
it can be constructed to do so by using $q^{1/\nu}$ instead of $q^{2-\eta}$ 
in Eq.(\ref{Hcf}).

\section{Conclusion}	\label{sDiscuss}

The analysis that is presented here is aimed at emphasizing the crucial
role of the type of boundary conditions on fluctuation--induced 
interactions. Unlike the case of extended objects at close separations,
where different types of boundary conditions all lead to the same 
form of interaction, we found that for objects at large separations
the type of boundary conditions determine the very form of the interaction,
and whether or not it is pairwise additive.

\acknowledgments

I am grateful to M. Kardar for invaluable discussions and comments. This research was 
supported in part by the National Science Foundation under Grants 
No. PHY94-07194 and DMR-98-05833.

\end{multicols}

\begin{references}

%\bibitem[*]{Add1}
%Present address

\bibitem{Casimir}
H.B.G. Casimir, Proc. Kon. Ned. Akad. Wetenschap B {\bf 51}, 793 (1948).

\bibitem{Dzya}
I.E. Dzyaloshinskii, E.M. Lifshitz, and L.P. Pitaevskii, Advan. Phys. {\bf 10},
165 (1961).

\bibitem{mostepa}
V.M. Mostepanenko and N.N. Trunov, {\em The Casimir Effect and Its Applications}
(Clarendon Press, Oxford, 1997).

\bibitem{Krech}
M. Krech, {\em The Casimir Effect in Critical Systems} (World Scientific,
Singapore, 1994).

\bibitem{KGRMP}
M. Kardar and R. Golestanian, Rev. Mod. Phys. {\bf 71}, 1233 (1999).

\bibitem{HaLiu1}
B.-Y. Ha and A.J. Liu, Phys. Rev. Lett. {\bf 79}, 1289 (1997).

\bibitem{LiK}
H. Li and M. Kardar, Phys. Rev. Lett. {\bf 67}, 3275 (1991); Phys. Rev. A
{\bf 46}, 6490 (1992).

\bibitem{HaLiu2}
B.-Y. Ha and A.J. Liu, Phys. Rev. Lett. {\bf 81}, 1011 (1998).

\bibitem{Pod}
R. Podgornik and V.A. Parsegian, Phys. Rev. Lett. {\bf 80}, 1560 (1998).

\bibitem{GGK}
R. Golestanian, M. Goulian, and M. Kardar, Europhys. Lett. {\bf 33}, 241 (1996);
Phys. Rev. E {\bf 54}, 6725 (1996).
 
\bibitem{dGF}
M.E. Fisher and P.-G. de Gennes, C. R. Acad. Sci. Paris B {\bf 287}, 207 (1978);
V. Privman and M.E. Fisher, Phys. Rev. B {\bf 30}, 322 (1984).

\bibitem{GBP}
M. Goulian, R. Bruinsma, and P. Pincus, Europhys. Lett. {\bf 22}, 145 (1993);
Erratum in Europhys. Lett. {\bf 23}, 155 (1993)

\bibitem{dG}
P.-G. de Gennes, C. R. Acad. Sci. Paris II {\bf 292}, 701 (1981).

\bibitem{BuEi}
T.W. Burkhardt and E. Eisenriegler, Phys. Rev. Lett. {\bf 74}, 3189 (1995).

\bibitem{betanu}
This interaction is in fact reported as $1/R^{2\beta/\nu}$ in Ref.\cite{BuEi}.
We have used the exponent identity $2\beta/\nu=d-2+\eta$ to write the power law
in such a way that its relation to the Gaussian case (where $\eta=0$) is manifest.

\end{references}
\end{document}